\documentstyle[aps,epsf,prd,floats,cite,preprint]{revtex}

\begin{document}

\preprint{\vbox{ \hbox{CERN-TH/99-357} \hbox{UCSD/PTH 99--21}
                 \hbox{hep-ph/9911461} \hbox{November 1999}
                 \hbox{} } }
\title{Charm Effects in the $\overline{\mbox{MS}}$ Bottom Quark Mass from 
$\Upsilon$ Mesons} 
\author{ A.~H.~Hoang${}^a$ and A V.~Manohar${}^b$}
\address{${}^a$ Theory Division, CERN,\\
   CH-1211 Geneva 23, Switzerland}
\address{${}^b$ Department of Physics, University of California at 
San Diego,\\
  9500 Gilman Drive, La Jolla, CA 92093-0319}
\maketitle
\begin{abstract}
We study the shift in the $\Upsilon$ mass due to a non-zero charm quark mass.
This shift affects the value of the $\overline{\rm MS}$ $b$-quark mass
extracted from the $\Upsilon$ system by about $-20$~MeV, due to an
incomplete cancellation of terms that are non-analytic in the charm
quark mass. The precise size of the shift depends on unknown higher order
corrections, and might have a considerable uncertainty if they are large.
\end{abstract}

\tighten

\begin{figure}[t] 
\begin{center}
\leavevmode
\epsfxsize=2.5
cm
\leavevmode
\epsffile[220 450 420 580]{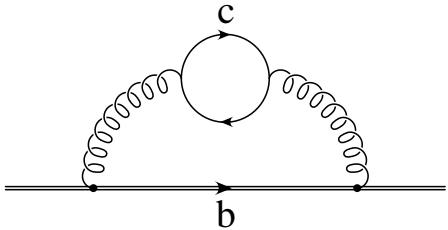}
%
%
\vskip  2.0cm
 \caption{Diagram contributing to the residual mass term of the $b$ quark
in HQET.\label{fig:twoloop} 
}
 \end{center}
\end{figure}

The bottom quark mass is an important parameter for the theoretical description
of $B$ meson decays and $b$ jet production cross sections in  collider
experiments. In continuum QCD, the most precise determinations of the bottom
quark mass parameter have been obtained from data on the spectrum and the
electronic partial widths of the $\Upsilon$ mesons.  Recently, a number of
$\overline{\mbox{MS}}$ bottom quark determinations have been carried out, which
were based on $\Upsilon$ meson sum rules at next-to-next-to-leading order
(NNLO) in the non-relativistic expansion, and which consistently eliminated all
linear sensitivity to small momenta~\cite{Melnikov1,Hoang1,Beneke1}. The latter
is mandatory  to reduce the systematic uncertainty in the bottom quark mass
below the typical hadronization scale $\Lambda_{\rm
QCD}$~\cite{Hoang2,Beneke2}.   The analyses mentioned above, however, treated
all quarks other than the $b$ quark as massless. This treatment is justified
for  light quarks that have masses much smaller than the inverse Bohr radius
$1/\langle r\rangle$ of the non-relativistic bottom--antibottom system  i.e. for
up, down and strange quarks, because in this case the theoretical expressions
describing the bottom--antibottom dynamics and  the conversion to the
$\overline{\mbox{MS}}$ bottom mass definition can be expanded in the light
quark masses. Like the contributions that are linearly sensitive to small
momenta, the terms linear (and non-analytic) in these light quark masses cancel
out in the analysis. At NLO in the non-relativistic expansion this can be seen
explicitly by considering the effects of a light virtual quark to the static
energy of a bottom--antibottom quark pair with spatial distance $r$,\footnote{
The dominant terms that are linearly sensitive to  low momenta in the
Schr\"odinger equation that describes the bottom--antibottom dynamics are all
contained in the  static energy.~\cite{Hoang2,Beneke2}}
\begin{eqnarray}\label{estatic}
E_{\rm stat} & = & 
2 M_b + V_{\rm stat}(r)
\,,
\end{eqnarray}  
where $M_b$ is the bottom quark pole mass and $V_{\rm stat}$ the potential
energy of the non-relativistic bottom--antibottom quark system. At order
$\alpha_s^2$ the correction coming from the finite mass of a light
quark $q$ to the pole mass contribution 
reads~\cite{Broadhurst1}
\begin{eqnarray}  
\delta M_b^q & = & 
\frac{4}{3}\,\frac{\alpha_s^2}{\pi^2}\,M_b\,\Delta\Big(\frac{m_q}{M_b}\Big)
\,, \label{gray}
\label{lightquarkmass}
\\[2mm] 
\Delta(r) & = &
\frac{\pi^2}{8}\,r - \frac{3}{4}\,r^2 + 
\frac{\pi^2}{8}\,r^3 - 
(\frac{1}{4}\,\ln^2r-\frac{13}{24}\,\ln r + 
 \frac{\pi^2}{24} + \frac{151}{288})\,r^4 
\nonumber
\\[1mm] & & -
\sum\limits_{n=3}^\infty\Big(2\,F(n)\,\ln r+F^\prime(n)\Big)\,r^{2n}
\,,
\\[2mm]
F(n) & \equiv & \frac{3\,(n-1)}{4\,n\,(n-2)\,(2n-1)\,(2n-3)}
\,,
\end{eqnarray}
where $\delta M^q_b$ is the shift in the $b$-quark pole mass keeping the
$b$-quark $\overline{\mbox{MS}}$ mass fixed, $m_q$ is the  mass of the
light quark, and $F^\prime(n)\equiv\frac{\partial}{\partial n} F(n)$.
 In the limit $m_q/M_b \to 0$, Eq.~(\ref{gray}) reduces to
\begin{equation}\label{eft}
\delta M_b^q  = {1\over6}\alpha_s^2 m_c.
\end{equation}
The shift $\delta M_b^q$ is non-analytic in the quark masses, and should be
regarded as being of the form $\delta M_b^q  = \alpha_s^2 \sqrt{m_c^2/M_b^2}
\left(M_b/6 \right)$, so that it is explicitly proportional to $M_b$, which
breaks the $b$-quark chiral symmetry. The limiting value Eq.~(\ref{eft}) can be
easily computed using heavy quark effective theory. At the scale $\mu=m_b$, one
matches to a theory in which the $b$-quark is treated as a static field, with
the residual mass term equal to zero, so that the propagator is $i/(k\cdot v)$.
At the lower scale $\mu=m_c$, one integrates out the charm quark. The matching
condition at this scale induces a residual mass term for the $b$-quark, which
is given by computing the graph in Fig.~\ref{fig:twoloop}

%
%

The shift in the bottom--antibottom quark potential energy due to a
light quark $q$ is 
\begin{eqnarray}
\delta V_{\rm stat}^q(r) & = &
-\frac{4}{3}\,\frac{\alpha_s}{r}\,
\bigg(\frac{\alpha_s}{6\,\pi}\bigg)\,\bigg\{\,
\ln\Big(\frac{m_q^2}{\mu^2}\Big) +
\int\limits_{4m_q^2}^\infty\frac{d\lambda^2}{\lambda^2}\,
R_{q\bar q}(m_q,\lambda)\,\exp(-\lambda\,r)
\,\bigg\}
\,,
\\[2mm]
R_{q\bar q}(m_q,\lambda) & = & 
\sqrt{1-\frac{4\,m_q^2}{\lambda^2}}\,
\bigg(\,1+\frac{2\,m_q^2}{\lambda^2}
\,\bigg)
\label{lightquarkpotential}
\end{eqnarray}
where $R_{q\bar q}$ is the
$q\bar q$ pair production cross section for the center-of-mass energy
$\lambda$ in $e^+e^-$--annihilation normalized to the massless cross
section. 
For $m_q \ll 1/\langle r\rangle \approx M_b\,\alpha_s$ the static
potential energy can be expanded in $m_q\,r$:
\begin{eqnarray}
\delta V_{\rm stat}^q(r) & \longrightarrow &
\frac{4}{3}\,\frac{\alpha_s}{r}\,
\bigg(\frac{\alpha_s}{3\,\pi}\bigg)\,\bigg\{\,
\ln(\mu\,r) + \gamma + \frac{5}{6} 
\,\bigg\} - 
\frac{1}{3}\,\alpha_s^2\,m_q + 
{\cal{O}}(\frac{\alpha_s^2}{\pi}\,m_q^2\,r)
\,,
\end{eqnarray}
and the contribution linear in $m_q$ cancels in the total static energy,
Eq.~(\ref{estatic}) between $M_b$ and $V(r)$.  The dominant light quark
correction is of order $(\alpha_s/\pi)^2 m_q^2/M_b$ and can be neglected for
all practical purposes.
 
On the other hand, for the case  that a light quark mass is comparable in size
to the inverse Bohr radius, i.e. for the charm quark, the full $m_q$-dependence
of $\delta V_{\rm stat}^q$ has to be taken into account. In this case the
cancellation of the linear charm mass term between the pole mass and potential
contributions to the static energy, Eq.~(\ref{lightquarkmass}), is incomplete. 
In this letter we show that this feature can lead to sizeable effects in the
$\overline{\mbox{MS}}$ bottom quark mass determination, compared to when the
charm quark is treated as massless.

There are two reasons why the incomplete cancellation of the linear charm mass
contribution can lead to a sizeable effect. First, the linear charm mass term
is not suppressed by a factor $\pi^2$, as one might expect from a contribution
coming from a loop integration (see Eq.~(\ref{lightquarkmass})). This  is
because the linear charm mass contribution represents a correction  generated
by a non-analytic linear dependence on infrared momenta.  A similar effect is
known in chiral perturbation theory where non-analytic $m_\pi^3$ corrections
from loop diagrams are also not suppressed by powers of $\pi$. Second, the
scale of the strong coupling of the linear charm mass term in
Eq.~(\ref{lightquarkmass}) is of order the charm mass rather than the bottom
quark mass.  This can be understood from the fact that the charm mass serves 
as an infrared cutoff for the non-analytic linear dependence on infrared momenta
just mentioned before. Thus, the linear charm mass term is generated at momenta
of order $m_c$. The effective field theory computation of
Fig.~\ref{fig:twoloop} also indicates that one should use $\alpha_s(m_c)$.
These arguments are supported by an explicit calculation of the order
$\alpha_s^3$ BLM corrections to the linear charm mass term in
Eq.(\ref{lightquarkmass}). For this calculation it is sufficient to consider
the static limit, i.e. we only need to determine the linear charm mass
corrections to the bottom quark self-energy due to the chromostatic Coulomb
field. The order $\alpha_s^3$ BLM linear charm mass contributions in the bottom
quark pole mass can be obtained from the formula 
\begin{eqnarray} \delta M_{b, stat}^c
& = & \frac{\alpha_s}{6\pi}\, \int\limits_{4 M_c^2}^\infty \frac{d
\lambda^2}{\lambda^2}\, R_{q\bar q}(m_c,\lambda)\, \int \frac{d^3
\mbox{\boldmath $p$}}{(2\pi)^3}\, \frac{2}{3}\,
\frac{4\,\pi\,\alpha_s}{\mbox{\boldmath $p$}^2+\lambda^2} \bigg[\,1 + 
2\,\frac{\alpha_s}{4\pi}\,\beta_0 \bigg(\ln\Big(\frac{\mu^2}{\mbox{\boldmath
$p$}^2}\Big)+\frac{5}{3}\bigg) \,\bigg] \,, 
\label{BLMcalculation}
\end{eqnarray}
where we assumed that $m_c$ is the charm pole mass, and $\beta_0=11-\frac{2}{3}
n_l$  for $n_l=3$ light quark flavors. The factor of two in front of the
$\beta_0$ term arises because the charm quark loop can be inserted on both
sides of the massless fermion loops. We have also included the chromostatic
self-energy contribution at order $\alpha_s^2$. The term linear in $m_c$
contained in Eq.~(\ref{BLMcalculation}) reads 
\begin{eqnarray} \delta M_{b,{\rm linear}}^c & = &
\frac{\alpha_s^2(\mu)}{6}\,m_c\,\bigg[\,
1+2\,\Big(\frac{\alpha_s}{4\pi}\Big)\,\beta_0\,
\bigg(\ln\Big(\frac{\mu^2}{m_c^2}\Big)-4\ln 2+\frac{14}{3}\bigg) \,\bigg] \,,
\end{eqnarray}
which corresponds to the BLM scale  $\mu_{\rm BLM}^{\rm pole}=0.388\,m_c$.
Obviously, the BLM calculation indicates a very low renormalization scale for
the strong coupling governing the linear charm quark mass contribution.

To illustrate the size of the corrections caused by the incomplete cancellation
of the linear charm quark terms let us examine the difference in the
$\overline{\mbox{MS}}$ bottom quark mass  $\overline m_b(\overline m_b)$,
determined from the mass of the $\Upsilon(1S)$ meson, $M_{\Upsilon(1S)} =
9.460$~GeV, for the two cases that that the finite charm mass effects are
either taken into account or neglected. For simplicity, we only consider an
extraction of the $\overline{\mbox{MS}}$ bottom quark mass at NLO in the
non-relativistic expansion. A more thorough analysis using full NNLO expressions
and including also a sum rule analysis based on data for all known $\Upsilon$
mesons will be carried out elsewhere~\cite{Hoang3}.

Including the effects of the charm quark mass from
Eq.~(\ref{lightquarkpotential}) properly in first order time-independent
perturbation theory, the  $\Upsilon(1S)$ meson mass at NLO in the
non-relativistic expansion reads~\cite{Titard1,Soto1}
($a_s\equiv\alpha_s^{(n_f=4)}(\mu)$) 
\begin{eqnarray}
M_{\Upsilon(1S)} & = & 2\,M_b\,\bigg\{\,1 \,-\,
 \frac{C_F^2\,a_s^2}{8}
\nonumber
\\[2mm] & &
\,-
\frac{C_F^2\,a_s^2}{8}\, 
\Big(\frac{a_s}{\pi}\Big)\,\bigg[\,
\beta_0\,\bigg( L + 1 \,\bigg) + \frac{a_1}{2} 
+\frac{2}{3}\bigg(\ln\Big(\frac{m_c}{\mu}\Big)+
 h\Big(\frac{2 m_c}{M_b C_F a_s}\Big)
\,\bigg)
\,\bigg]
\,\bigg\}
\,,
\label{M1Sdef}
\end{eqnarray}
where~\cite{Soto1}
\begin{eqnarray}
h(x) & \equiv & 
-\frac{11}{6} - 2\,x^2 + \frac{3\,x\,\pi }{4} + 
  x^3\,\pi  +
\left\{
  \begin{array}{c@{\quad:\quad}l}
  \frac{( 2 - x^2 - 4\,x^4 )}{2\,\sqrt{x^2-1}}\,
     \tan^{-1}\left({\sqrt{x^2-1}}\right) & x>1 \\[2mm]
  \frac{( 2 - x^2 - 4\,x^4 )}{2} & x=1 \\[2mm]
  \frac{( 2 - x^2 - 4\,x^4 )}{4\,\sqrt{1-x^2}}\,
     \ln\left(\frac{1+\sqrt{1-x^2}}{1-\sqrt{1-x^2}}\right) & x<1
  \end{array}
\right.
\,,
\\[2mm]
L & = & \ln\Big(\frac{\mu}{C_F\,a_s\,M_b}\Big)
\,,
\end{eqnarray}
and $a_1 = \frac{31}{3} - \frac{10}{9} n_l$  for $n_l=3$ massless quark
flavors~\cite{Fischler1,Billoire1}. For $x\to 0$ the function $h$ has the
limiting behavior $h(x\to 0)=-11/6+\ln(2/x)+3\pi x/4+{\cal{O}}(x^2)$.  Using
the upsilon expansion up to order $\epsilon^2$~\cite{Hoang4} and  the
expressions given in  Eqs.~(\ref{lightquarkmass}) and (\ref{M1Sdef}) we arrive
at the following formula for the shift in the $\overline{\mbox{MS}}$ bottom
quark mass ($C_F=4/3$):
\begin{eqnarray}
\Delta\overline m_b & = &
\frac{M_{\Upsilon(1S)}}{2}\,
\bigg\{\,
\frac{C_F^2\,\alpha_s^2}{8}\,\Big(\frac{\alpha_s}{\pi}\Big)\,
\frac{2}{3}\, 
\bigg[\,
h\Big(\frac{4\,m_c}{M_{\Upsilon(1S)}\,C_F\,\alpha_s}\Big)
+\frac{11}{6}
-\ln\Big(\frac{M_{\Upsilon(1S)}\,C_F\,\alpha_s}{2\,m_c}\Big)
\,\bigg]
\nonumber\\[2mm] & & \hspace{1cm}
-\frac{1}{3}\,\alpha_s^2\,\frac{m_c}{M_{\Upsilon(1S)}}\label{shift}
\,\bigg\}
\,,
\label{deltam}
\end{eqnarray}
where we have taken into account only the term linear in the charm quark mass
from Eq.~(\ref{lightquarkmass}).\footnote{It is sufficient to only include the
term linear in $m_c$ in Eq.~(\ref{lightquarkmass}), since the higher order
terms are suppressed by powers of $m_c/M_b$ which is small. Higher order terms
in Eq.~(\ref{M1Sdef}) depend on powers of $2 M_c/(M_b C_F a_s)$, which is not
small, and it is necessary to include the full functional dependence in
$h(x)$.} 
The scale of the strong coupling contained in the first line of
Eq.~(\ref{deltam}) is of order the inverse Bohr radius. We identify
this scale with the one of the linear charm mass term.
In deriving Eq.~(\ref{shift}), the terms $-11/6+\ln(2/x)$ in $h(x\to
0)$ have been absorbed into  Eq.~(\ref{M1Sdef}), with the replacement $n_l \to
n_l+1=4$. This gives the usual relation between the $1S$ and $\overline{\rm
MS}$ masses neglecting the charm quark mass, so these terms in $h(x)$ do not
contribute to the shift  $\Delta\overline m_b$.

\begin{figure}[t] 
\begin{center}
\leavevmode
\epsfxsize=5cm
\leavevmode
\epsffile[220 580 420 710]{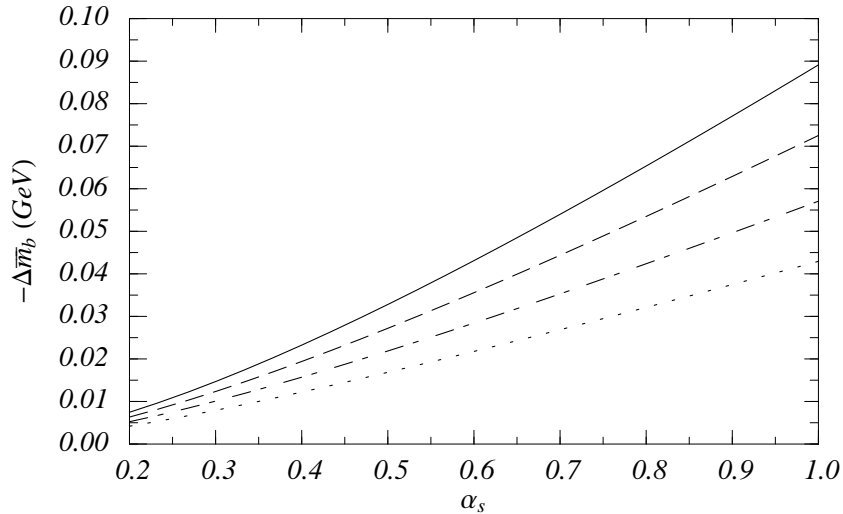}
%
%
\vskip  3.6cm
 \caption{\label{figshift} 
The function $-\Delta\overline m_b$ as a function of $\alpha_s$. The
solid, dashed, dash-dotted and dotted lines correspond to the choices
$m_c=1.7, 1.5, 1.3$ and $1.1$~GeV, respectively, for the charm quark
mass. 
}
 \end{center}
\end{figure}

In Fig.~\ref{figshift} we have displayed
$-\Delta\overline m_b$ as a function of $\alpha_s$.  $\alpha_s$ has to be
evaluated at a scale of order the charm mass as discussed above. The solid,
dashed, dash-dotted and dotted lines correspond to the choices $m_c=1.7, 1.5,
1.3$ and $1.1$~GeV, respectively, for the charm quark mass. For
$\alpha_s\approx0.4$, which corresponds to a choice of the renormalization
scale equal to the charm quark mass, we find that the shift is between $-10$ and
$-20$~MeV. If a smaller renormalization scale is assumed, the shift can amount to
more than $-50$~MeV for larger choices of the charm quark mass. The spread in the
curves displayed in Fig.~\ref{figshift} shows that the shift in the
$\overline{\mbox{MS}}$ bottom quark mass can be of order several tens of MeV. The
exact value, however, contains a considerable uncertainty, which is amplified
by the large value of the strong coupling governing the $\Delta\overline m_b$.
A complete NNLO analysis for the $\overline{\mbox{MS}}$ bottom quark extraction
from $\Upsilon$ sum rules will be indispensable to accurately determine the
effect of a nonzero charm mass. However, if the scale governing the strong
coupling constant in $\Delta\overline m_b$ is indeed as low as the BLM scale
estimate carried out above indicates, a considerable uncertainty might 
persist. For a moderate choice of the strong coupling in $\Delta\overline m_b$
between 0.4 and 0.6 the charm mass shift is smaller than the uncertainties of
60--80~MeV in the value of  $\overline m_b(\overline m_b)$ obtained in recent
NNLO analyses of the $\Upsilon$ sum rules~\cite{Melnikov1,Hoang1,Beneke1},
where charm mass effects have been neglected. However, the inclusion of the
charm mass effects will be essential for a future extraction of $\overline
m_b(\overline m_b)$ at NNNLO.

AH would like thank the members of the UCSD high energy theory group for their
hospitality, and for the pleasant time at the UCSD Physics Department,
where this work was finalized.  AH is supported in part by the EU
Fourth Framework Program ``Training and Mobility of Researchers'',
Network ``Quantum Chromodynamics and Deep Structure of Elementary
Particles'', contract FMRX-CT98-0194 (DG12-MIHT).  AM is supported in
part by the U.S.~Department of Energy under
contract~DOE~DE-FG03-97ER40546.

\sloppy
\raggedright
\def\app#1#2#3{{\it Act. Phys. Pol. }{\bf B #1} (#2) #3}
\def\apa#1#2#3{{\it Act. Phys. Austr.}{\bf #1} (#2) #3}
\def\lhc{Proc. LHC Workshop, CERN 90-10}
\def\npb#1#2#3{{\it Nucl. Phys. }{\bf B #1} (#2) #3}
\def\nP#1#2#3{{\it Nucl. Phys. }{\bf #1} (#2) #3}
\def\plb#1#2#3{{\it Phys. Lett. }{\bf B #1} (#2) #3}
\def\prd#1#2#3{{\it Phys. Rev. }{\bf D #1} (#2) #3}
\def\pra#1#2#3{{\it Phys. Rev. }{\bf A #1} (#2) #3}
\def\pR#1#2#3{{\it Phys. Rev. }{\bf #1} (#2) #3}
\def\prl#1#2#3{{\it Phys. Rev. Lett. }{\bf #1} (#2) #3}
\def\prc#1#2#3{{\it Phys. Reports }{\bf #1} (#2) #3}
\def\cpc#1#2#3{{\it Comp. Phys. Commun. }{\bf #1} (#2) #3}
\def\nim#1#2#3{{\it Nucl. Inst. Meth. }{\bf #1} (#2) #3}
\def\pr#1#2#3{{\it Phys. Reports }{\bf #1} (#2) #3}
\def\sovnp#1#2#3{{\it Sov. J. Nucl. Phys. }{\bf #1} (#2) #3}
\def\sovpJ#1#2#3{{\it Sov. Phys. LETP Lett. }{\bf #1} (#2) #3}
\def\jl#1#2#3{{\it JETP Lett. }{\bf #1} (#2) #3}
\def\jet#1#2#3{{\it JETP Lett. }{\bf #1} (#2) #3}
\def\zpc#1#2#3{{\it Z. Phys. }{\bf C #1} (#2) #3}
\def\ptp#1#2#3{{\it Prog.~Theor.~Phys.~}{\bf #1} (#2) #3}
\def\nca#1#2#3{{\it Nuovo~Cim.~}{\bf #1A} (#2) #3}
\def\ap#1#2#3{{\it Ann. Phys. }{\bf #1} (#2) #3}
\def\hpa#1#2#3{{\it Helv. Phys. Acta }{\bf #1} (#2) #3}
\def\ijmpA#1#2#3{{\it Int. J. Mod. Phys. }{\bf A #1} (#2) #3}
\def\ZETF#1#2#3{{\it Pis'ma Zh. Eksp. Teor. Fiz. }{\bf #1} (#2) #3}
\def\jmp#1#2#3{{\it J. Math. Phys. }{\bf #1} (#2) #3}
\def\yf#1#2#3{{\it Yad. Fiz. }{\bf #1} (#2) #3}
\def\aspn#1#2#3{{\it Arch. Sci. Phys. Nat. }{\bf #1} (#2) #3}


\begin{thebibliography}{99}
%
%
\bibitem{Melnikov1}
K. Melnikov and A. Yelkhovky, \prd{59}{1999}{114009}
%
\bibitem{Hoang1} 
A. H. Hoang, to appear in {\it Phys. Rev. }{\bf D}, 
hep-ph/9905550.
%
\bibitem{Beneke1}
M. Beneke and A. Signer, hep-ph/9906475.
%
\bibitem{Hoang2}
A. H. Hoang, M. C. Smith, T. Stelzer and S. S. Willenbrock,
\prd{59}{1999}{114014}.
%
\bibitem{Beneke2}
M. Beneke, \plb{434}{1998}{115}.
%
\bibitem{Broadhurst1}
N. Gray, D. J. Broadhurst, W. Grafe and K. Schilcher,
\zpc{48}{1990}{673}.
%
\bibitem{MW}
A.V.~Manohar and M.B.~Wise, unpublished.
%
\bibitem{Hoang3}
A. H. Hoang and M. Melles, in preparation.
%
\bibitem{Titard1}
S. Titard and F. J. Yndurain, \prd{49}{1994}{6007}.
%
\bibitem{Soto1}
D. Eiras and J. Soto,  hep-ph/9905543. 
%
\bibitem{Fischler1}
W. Fischler, \npb{129}{1977}{157}.
%
\bibitem{Billoire1}
A. Billoire, \plb{92}{1980}{343}.
%
\bibitem{Hoang4}
A. H. Hoang, Z. Ligeti and A. V. Manohar, \prl{82}{1999}{277},\\
\prd{59}{1999}{074017}.
%
%
\end{thebibliography}
\end{document}